\documentclass[aps,preprint,preprintnumbers,nofootinbib,showpacs]{revtex4}
\usepackage{graphicx,color,amsmath}
\begin{document}
\title{Constraints on the new particle in $\Sigma^+\to p\mu^+\mu^-$}
\author{C. Q. Geng and Y. K. Hsiao} \affiliation{Department of Physics,
National Tsing Hua University, Hsinchu, Taiwan 300}
\date{\today}
\begin{abstract}
The HyperCP collaboration has presented the branching ratio of
$\Sigma^+\to p\mu^+\mu^-$ to be $(8.6^{+6.6}_{-5.4}\pm 5.5)\times
10^{-8}$ and suggested a new boson $P^0$ with a mass of $214.3\pm
0.5$ MeV to induce the flavor changing transition of $s\to
d\mu^+\mu^-$. We demonstrate that to explain the data, the new
particle cannot be a scalar but pseudoscalar based on the direct
constraints from $K^+\to\pi^+\mu^+\mu^-$ and $K_L\to\mu^+\mu^-$,
respectively. Moreover, we determine that the decay width of the
pseudoscalar should be in the range of $10^{-7}$ MeV with the
lifetime of $10^{-14}$ sec.
\end{abstract}


\maketitle

\begin{figure}[htb]
\centering
\includegraphics[width=2.1in]{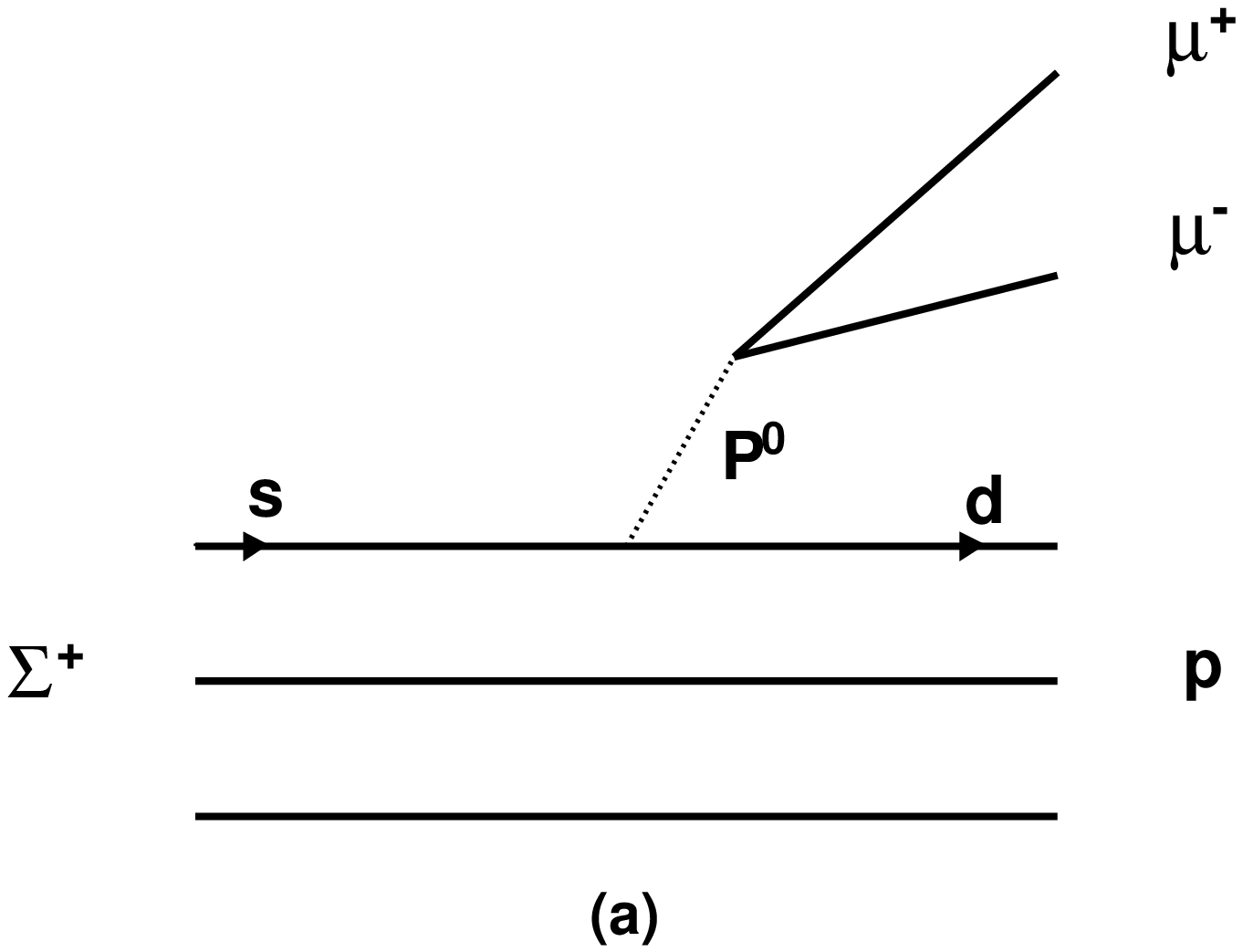}
\includegraphics[width=2.1in]{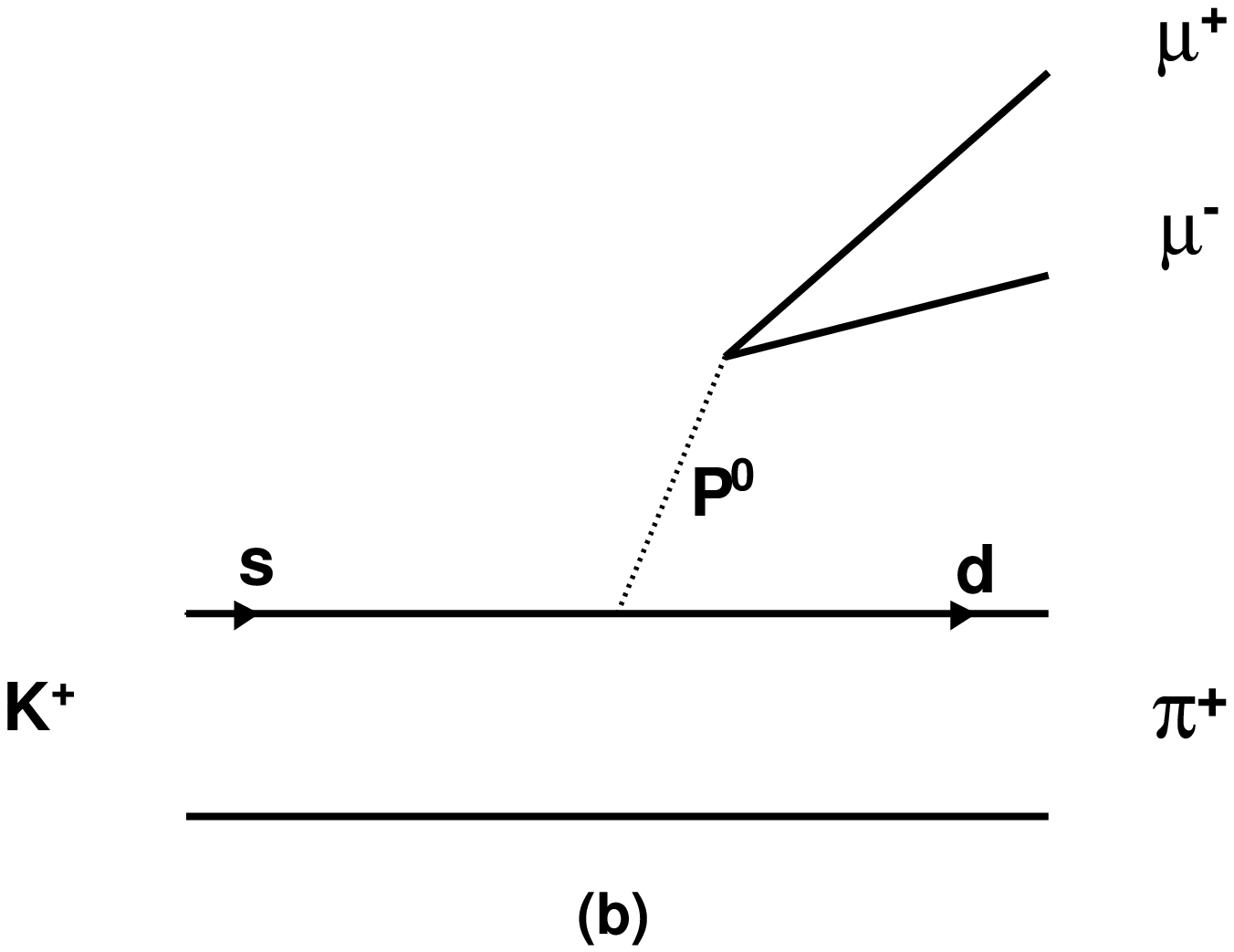}
\includegraphics[width=2.1in]{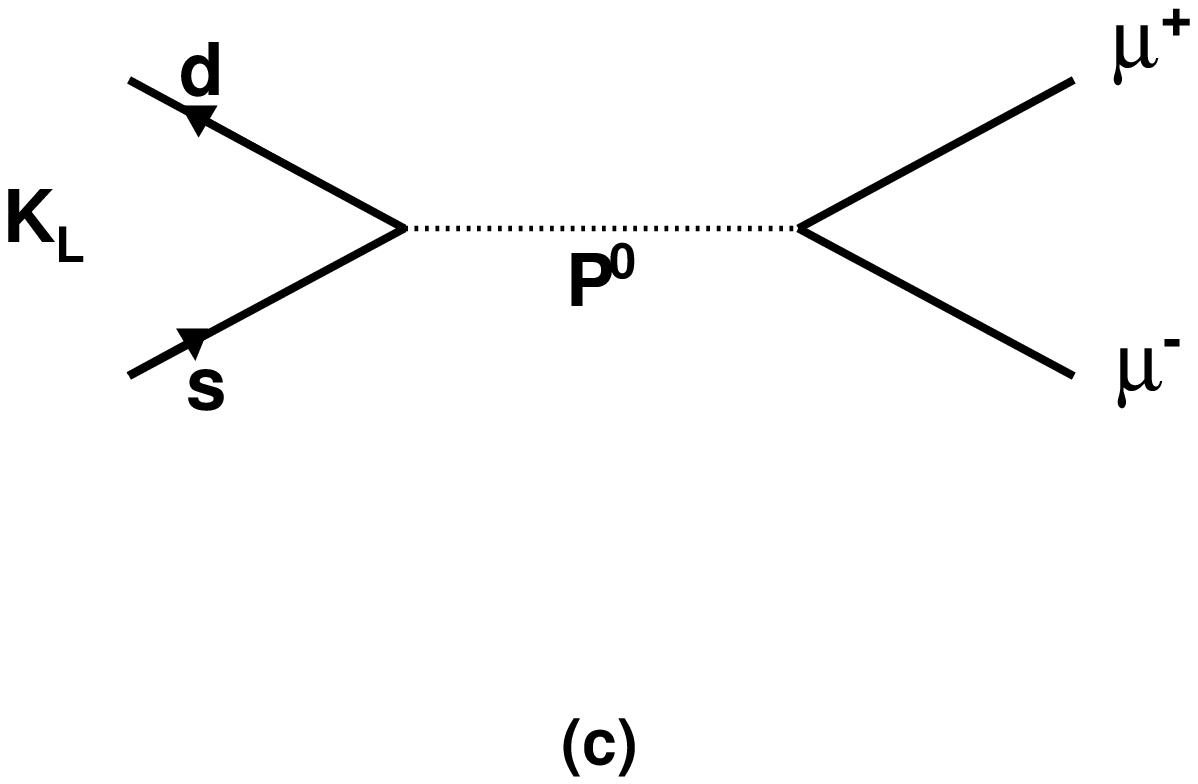}
\caption{Diagrams for (a) $\Sigma^+ \to pP^0, P^0\to \mu^+\mu^-$,
(b) $K^+ \to \pi^+P^0, P^0\to \mu^+\mu^-$ and
(c) $K_L \to P^0, P^0\to \mu^+\mu^-$.}\label{fig}
\end{figure}

According to the report of the HyperCP collaboration \cite{Park},
the observation of three events for the decay $\Sigma^+\to
p\mu^+\mu^-$ reveals the possibilities of new physics, as the
branching ratio Br($\Sigma^+\to p\mu^+\mu^-)=(8.6^{+6.6}_{-5.4}\pm
5.5)\times 10^{-8}$ is claimed to be larger than the prediction
within the Standard Model \cite{Park,Bergstrom,He}. The analysis
in Ref. \cite{Park} has found an unexpectedly narrow dimuon
distribution, which can not be explained by the form factors used
to deform the phase space due to their mildly momentum dependences
\cite{Bergstrom, He}. The plausible explanation can be the
threshold effect which is induced as
$m_{\mu^+\mu^-}=(p_{\mu^+}+p_{\mu^-})$ is approaching to the pole
of some unknown intermediate boson, suggesting a two-body
$\Sigma^+\to pP^0,\,P^0\to\mu^+\mu^-$ decay shown in Fig.
\ref{fig}a, with the $P^0$ mass being $m_{P^0}=214.3\pm
0.5\;\text{MeV}$ \cite{Park} and the branching ratio \cite{Park}
\begin{eqnarray}\label{exp31}
Br(\Sigma^+\to pP^0,\,P^0\to\mu^+\mu^-)=(3.1^{+2.4}_{-1.9}\pm
1.5)\times 10^{-8}.
\end{eqnarray}
If the effect is true, the flavor-changing neutral current (FCNC)
of the $s\to d$ transition is discovered at tree level. Clearly,
the most important task is to check the reality of the experiment.
Note that the observed events are only three and the physical
properties of this unknown particle remain ambiguous.
Nevertheless, the investigations can proceed via the decays of
$K^+\to\pi^+\mu^+\mu^-$ and $K_L\to\mu^+\mu^-$ since they share
the same effective four-fermion interaction at quark level as
shown in Fig. \ref{fig}. In this paper, we will explore the
constraints on the new particle suggested by the HyperCP
collaboration by relating the three decay modes.

We start with the general effective four-fermion interaction for
$s\to d\mu^+\mu^-$ in Fig. \ref{fig} by including all possible
scalar-type currents, given by
\begin{eqnarray}\label{HNP}
{\cal L}_{NP}=
 \frac{\lambda_{ij}}{q^2-m_{P^0}^2+im_{P^0}\Gamma_{P^0}}\bar d\Gamma_is\bar u\Gamma_jv+H.C.\;,
\end{eqnarray}
where $q=p_{\mu^+}+p_{\mu^-}$, $\Gamma_{P^0}$ is the decay width,
$u$ ($v$) denotes the $\mu^-$ ($\mu^+$) spinor and $\lambda_{ij}$
are the combined coupling constants with $i,j=S$ and $P$ for
$\Gamma_{i,j}=1$ and $\gamma_5$, representing scalar and
pseudoscalar currents, respectively. We stress the necessity of
the decay width $\Gamma_{P^0}$ in Eq. (\ref{HNP}) for the
threshold enhancement around the pole near
$m_{\mu^+\mu^-}=214.3\;\text{MeV}$. In Eq. (\ref{HNP}), there are
four kinds of couplings through $S\otimes S$, $P\otimes P$,
$S\otimes P$ and $P\otimes S$ currents. Note that the latter two
are parity-odd terms with the physical states being the mixtures
of scalar and pseudoscalar \cite{GN}. Moreover, the last one could
also violate CP symmetry through the longitudinal muon
polarization in $K_L\to \mu^+\mu^-$ \cite{GN}. However, we shall
not consider CP violation in the present paper.

We now apply ${\cal L}_{NP}$ in Eq. (\ref{HNP}) to the decay of
$\Sigma^+\to pP^0, P^0\to\mu^+\mu^-$. The amplitude is found to be
\begin{eqnarray}\label{sigma}
{\cal A}_{\Sigma^+}\equiv{\cal A}(\Sigma^+\to pP^0,
P^0\to\mu^+\mu^-)=
\frac{\lambda_{ij}}{q^2-m_{P^0}^2+im_{P^0}\Gamma_{P^0}}\langle
p|\bar d\Gamma_is|\Sigma^+\rangle \bar u\Gamma_jv\;.
\end{eqnarray}
To evaluate the amplitude, we parametrize
\begin{eqnarray}
\langle p|\bar ds|\Sigma^+\rangle=f_S\bar u_p u_\Sigma\,,\;\;
\langle p|\bar d\gamma_5s|\Sigma^+\rangle=g_P\bar
u_p\gamma_5u_\Sigma,
\end{eqnarray}
where the form factors are given by \cite{betadecay1}
\begin{eqnarray}
f_S=f_1(q^2)\frac{m_\Sigma-m_p}{m_s-m_d}\;,\;\;\;g_P=g_1(q^2)\frac{m_\Sigma+m_p}{m_s+m_d}\;,
\end{eqnarray}
with \cite{betadecay1,betadecay2}
\begin{eqnarray}
&&f_1(q^2)=\frac{f_1(0)}{(1-\frac{q^2}{m_V^2})^2}\,,\;\;\;g_1(q^2)=\frac{g_1(0)}{(1-\frac{q^2}{m_A^2})^2}\;,
\nonumber\\
 &&f_1(0)=-1.0\,,\ g_1(0)=0.35\,,\
m_V=0.97\ GeV\,,\  m_A=1.25\; GeV.
\end{eqnarray}
It is noted that the momentum dependences are expressed as the
double-pole expansions.

To test possibilities of new physics from $\Sigma^+\to
pP^0,P^0\to\mu^+\mu^-$, we study the decays of
$K^+\to\pi^+\mu^+\mu^-$ and $K_L\to\mu^+\mu^-$ due to ${\cal
L}_{NP}$ in Eq. (\ref{HNP}). The amplitudes of
$K^+\to\pi^+P^0,P^0\to\mu^+\mu^-$ and $K_L\to
P^0,P^0\to\mu^+\mu^-$ are given by
\begin{eqnarray}\label{K3}
{\cal A}_{K^+}&=&
\frac{\lambda_{ij}}{q^2-m_{P^0}^2+im_{P^0}\Gamma_{P^0}}
\langle\pi^+|\bar d\Gamma_is|K^+\rangle \bar u\Gamma_jv\;,\nonumber\\
{\cal
A}_{K_L}&=&\frac{\lambda_{ij}}{q^2-m_{P^0}^2+im_{P^0}\Gamma_{P^0}}
\bigg[\langle 0|\bar d\Gamma_i s|K_L\rangle+\langle 0|\bar
s\Gamma_i d |K_L\rangle\bigg]\bar u\Gamma_jv\;,
\end{eqnarray}
respectively. Here, we have defined ${\cal A}_{K^+}\equiv{\cal
A}(K^+\to\pi^+P^0,P^0\to\mu^+\mu^-)$ and ${\cal
A}_{K_L}\equiv{\cal A}(K_L\to P^0,P^0\to\mu^+\mu^-)$. It is noted
that there are no contributions from $\langle\pi^+|\bar
d\gamma_5s|K^+\rangle$ and $\langle 0|\bar d s|K_L\rangle+\langle
0|\bar sd |K_L\rangle$ due to the parity conservation in strong
interaction. Therefore, $K^+\to\pi^+\mu^+\mu^-$ can only be used
to constrain the couplings of $S \otimes S(P)$, while
$K_L\to\mu^+\mu^-$ to those of $P\otimes S(P)$. The matrix
elements in Eq. (\ref{K3}) by means of equation of motion are
found to be
\begin{eqnarray}
\langle\pi^+|\bar ds|K^+\rangle&=&\frac{m^2_K-m^2_\pi}{m_s-m_d}f_+\;,\nonumber\\
\langle 0|\bar d\gamma_5 s|K_L\rangle+\langle 0|\bar s\gamma_5 d|K_L\rangle&=&i\sqrt 2 f_K\frac{m_K^2}{m_s+m_d}\;,
\end{eqnarray}
where $f_+\simeq 1$ and $f_K=160$ MeV \cite{pdg}.

To proceed, we first concentrate on $S \otimes S(P)$ couplings.
The experimental measurement for the decay of $K^+\to
\pi^+\mu^+\mu^-$ is \cite{pdg}
\begin{eqnarray}\label{kpimumu}
Br(K^+\to\pi^+\mu^+\mu^-)&=&(8.1\pm 1.4)\times 10^{-8}\,.
\end{eqnarray}
It has been domenstrated that the dominate contribution for the
decay is from the one-photon exchange in the Standard Model, which
can be referred, such as in Ref. \cite{ChPT}. Since the partial
branching ratio of the three-body decay is proportional to $|{\cal
A}|^2/m^3\cdot\tau$, where $|{\cal A}|^2$ is the squared amplitude
and $m$ ($\tau$) is the mass (lifetime) of the mother particle.
For the $S\otimes S\;(P)$ currents, from Eqs.
(\ref{sigma})-(\ref{K3}) we have
\begin{eqnarray}
|{\cal A}_{K^+}|^2/|{\cal A}_{\Sigma^+}|^2&\simeq& 0.25,
\end{eqnarray}
$(1/m_{K}^3)/(1/m_{\Sigma}^3)\simeq 14$ and
$\tau_{K^+}/\tau_{\Sigma^+}=1.5 \times 10^2$. After integrating
the phase space, we find that
\begin{eqnarray}
{Br(K^+\to\pi^+P^0,P^0\to\mu^+\mu^-)\over
Br(\Sigma^+\to pP^0,P^0\to\mu^+\mu^-)}&\sim &
O(10^2)\ (O(10^3))
\label{BRKS}
\end{eqnarray}
for the $S\otimes S\;(P)$ currents. Note that the estimation of
the ratio in Eq. (\ref{BRKS}) is independent of the property of
the new particle. When $Br(\Sigma^+\to pP^0,P^0\to\mu^+\mu^-)$ is
in the range of $O(10^{-8})$, in any case,
$Br(K^+\to\pi^+P^0,P^0\to\mu^+\mu^-)$ should be of
$O(10^{-6}-10^{-5})$ if the interaction is $S\otimes S\;(P)$,
which is clearly out of the limitation in Eq. (\ref{kpimumu}). As
a result, the tree level flavor-changing $s\to d\mu^+\mu^-$
transition resulting from the new physics of $S\otimes S\;(P)$
currents is unambiguously ruled out based on the data of
$K^+\to\pi^+\mu^+\mu^-$.

We now turn to the new physics from $P\otimes P$ and $P\otimes S$
currents. Currently, the experimental measurement on
$K_L\to\mu^+\mu^-$ is \cite{pdg}
\begin{eqnarray}\label{KL}
Br(K_L\to\mu^+\mu^-)=(6.87\pm 0.12)\times 10^{-9}\;,
\end{eqnarray}
which is almost saturated by the absorptive (imaginary) part,
dominated by the measured mode of $K_L\to\gamma\gamma$ with
$Br(K_L\to\gamma\gamma)=(5.56\pm 0.06)\times 10^{-4}$ \cite{pdg},
i.e., $Br_{abs}(K_L\to\mu^+\mu^-)=(6.66\pm 0.07)\times 10^{-9}$.
However, there is still a possibility of the cancellation among
the short-distance amplitude and the real part of the
long-distance part \cite{GN2,GH}. Nevertheless, it is believed
that the new physics contribution to the decay branching ratio
cannot excess of $O(10^{-9})$ \cite{GN2}. To be conservative, we
shall use
\begin{eqnarray}\label{KLA}
Br_{K_L}\equiv Br(K_L\to P^0,P^0\to\mu^+\mu^-)\leq 10^{-9}
\end{eqnarray}
as our working assumption to constrain the new physics.

Beginning with a rough estimate, if we assume that
$\Gamma_{P^0}\simeq 1$ MeV, we find that the ratio of $Br_{K_L}$
and $Br_{\Sigma^+}\equiv Br(\Sigma^+\to pP^0,P^0\to\mu^+\mu^-)$ is
around $10^{5}$ ($10^{6}$) for $P\otimes P(S)$ such that
$Br_{\Sigma^+}$ would be of order $10^{-13}$ ($10^{-14}$), which
is not consistent with the data in Eq. (\ref{exp31}) when relating
it to that of Eq. (\ref{KLA}). Nonetheless, if we resort the large
pole effect with a small $\Gamma_{P^0}$, there still remains a
possibility to make $\Sigma^+\to p\mu^+\mu^-$ in the order of
$10^{-8}$. It also explains why the constraint from
$K^+\to\pi^+\mu^+\mu^-$ is strong enough, so that there is no room
for new physics, since $K^+\to\pi^+\mu^+\mu^-$ and $\Sigma^+\to
p\mu^+\mu^-$ are both three-body decays and the pole effect is at
the same place. For $Br(K_L\to P^0, P^0\to\mu^+\mu^-)$ in the
range of $(0.3-10)\times 10^{-10}$ with the $P\otimes P$-type
current, and the allowed coupling constants and the decay width
are demonstrated in Table \ref{table}. While constraints on the
coupling constants are mainly from different open windows of
$Br(K_L\to P^0, P^0\to\mu^+\mu^-)$, the sets of $\Gamma_{P^0}$'s
are as small as possible to enhance $Br(\Sigma^+\to p P^0,P^0\to
\mu^+\mu^-)$ to match the experimental value  in Eq.
(\ref{exp31}), where the error has been taken as the larger one
between $\sigma_+$ and $\sigma_-$. It is obvious that the pole
effect plays the most important role to coincide with both data of
$K_L\to \mu^+\mu^-$ and $\Sigma^+\to p P^0, P^0\to\mu^+\mu^-$. As
a consequence, the coupling constant $\lambda_{PP}$ is in the
order of $10^{-13}$ while the decay width $\Gamma_{P^0}$ is in the
range of $10^{-7}\,\text{MeV}$, translated as the lifetime of
$\tau_{P^0}\simeq 10^{-14}\,\text{sec}$. Inasmuch as it is also
possible from the $P\otimes S$ current though it would induce CP
violation \cite{GN}, the window as well is opened as the same as
that of the $P\otimes P$ current, and we obtain that
$\lambda_{PS}$ around $10^{-13}$ with $\Gamma_{P^0}$ in
$10^{-8}\,\text{MeV}$, which is one order of magnitude smaller
than that of the $P\otimes P$ current.
\begin{table}[htb]
\footnotesize
\caption{{ \sl The coupling constant and decay width
of $P^0$ for $P\otimes P$ and $P\otimes S$
currents.}}\label{table}
\begin{tabular}{|c||c|c||c|c|}
\hline
\raisebox{-1ex}{$Br(K_L\to P^0,P^0$}&\multicolumn{2}{c}{\bf $P\otimes P$}\vline\,\vline&\multicolumn{2}{c}{\bf $P\otimes S$}\vline\\
\cline{2-5}
\raisebox{1ex}{$\to\mu^+\mu^-)$$(10^{-10}$)}  &   $|\lambda_{PP}|$ ($10^{-13}$)   &   $\Gamma_{P^0}$ ($10^{-7}$ MeV)  &   $|\lambda_{PS}|$ ($10^{-13}$)   &   $\Gamma_{P^0}$($10^{-9}$ MeV)   \\ \hline
0.3-0.6 &$  0.81^{+0.13}_{-0.15}    $&$ 0.92^{+11.75}_{-\;\;0.60}   $&$ 0.90^{+0.14}_{-0.16}    $&$ 3.10^{+39.57}_{-\;\;2.01}   $\\ \hline
0.6-1.0 &$  1.08^{+0.13}_{-0.15}    $&$ 1.64^{+19.47}_{-\;\;0.99}   $&$ 1.20^{+0.14}_{-0.16}    $&$ 5.51^{+65.63}_{-\;\;3.34}   $\\ \hline
1.0-1.5 &$  1.35^{+0.13}_{-0.14}    $&$ 2.56^{+29.14}_{-\;\;1.49}   $&$ 1.50^{+0.14}_{-0.16}    $&$ 8.61^{+98.09}_{-\;\;4.99}   $\\ \hline
1.5-2.0 &$  1.60^{+0.11}_{-0.12}    $&$ 3.58^{+38.62}_{-\;\,1.97}   $&$ 1.77^{+0.12}_{-0.13}    $&$ 12.05^{+130.18}_{-\;\;\;\,6.62} $\\ \hline
2.0-10  &$  2.97^{+0.86}_{-1.27}    $&$ 12.3^{+198.8}_{-\;\;10.2}   $&$ 3.28^{+0.95}_{-1.39}    $&$ 21.7^{+689.7}_{-\;\;14.5}   $\\ \hline
\end{tabular}
\end{table}

Among theoretical models, the proposed pseudoscalar particle $P^0$
is not likely the axion for its mass being much heavier than the
allowed values \cite{axion}. However, it cannot be a leptoquark
either since it could not lead to the pole effect. However, the
sgoldstino of the supersymmetric model is still allowed as the
range of the mass is consistent with the experimental one
\cite{sgoldtino}. Clearly, more efforts of both theory and
experiment are needed.

In sum, we have found that as an intermediate boson to the decay
of $\Sigma^+\to pP^0,P^0\to\mu^+\mu^-$, $P^0$ can be induced from
the $P\otimes P$ or $P\otimes S$-type current, which is testified
with $K_L\to \mu^+\mu^-$, whereas $S\otimes P$ and $S\otimes
S$-type currents have been proven to be impossible via the decay
of $K^+\to\pi^+\mu^+\mu^-$. Moreover, the analysis suggests that
in the window of $Br(K_L\to \mu^+\mu^-)=(0.3-0.6)\times 10^{-10}$,
while $Br(\Sigma^+\to pP^0,\,P^0\to\mu^+\mu^-)=(3.1\pm 2.8)\times
10^{-8}$ with $m_{P^0}=214.3\;\text{MeV}$, the decay width and
lifetime are $(0.92^{+11.75}_{-\;\;0.60})\times 10^{-7}$ MeV and
$(7.2^{+13.2}_{-\;\;2.2})\times 10^{-15}$ sec, respectively, with
the coupling constants $\lambda_{PP}=(0.81^{+0.13}_{-0.15})\times
10^{-13}$. Finally, we remark that our analysis can be generalized
to vector and axial-vector currents.

Note added: Before we presented the paper, there were two similar
papers by He {\it et al.} \cite{He2}, Deshpande {\it et al.}
\cite{Deshpande} and Gorbunov {\it et al.} \cite{sgoldtino2} in arXiv. Here, we make some comparisons as follows:\\
1. While Refs. \cite{He2,Deshpande}  begin with the Lagrangian
coupled to the new particle, we consider the effective four-quark
interaction at quark level, which leads to $\Sigma^+\to
p\mu^+\mu^-$. The propagator in the interaction includes the decay
width to avoid divergence due to the pole. As a result, we can
give stronger constraints on the decay width. We point out that in
the same effective four-quark interaction for $S\otimes S(P)$
currents, the deviation between $\Sigma^+\to p\mu^+\mu^-$ and
$K^+\to \pi^+\mu^+\mu^-$ is only in $|{\cal
A}|^2/m_{\Sigma(K)}^3\cdot \tau_{\Sigma(K)}$ with the phase space,
given a ratio of $Br_{K^+}/Br_{\Sigma^+}\simeq 10^{3}$, such that
we completely rule out any possibilities from a scalar coupling of
the $s\to d$ transition, which is consistent with
Ref. \cite{He2} while Ref. \cite{Deshpande} leaves ambiguity.\\
2. We constrain the couplings of $s\to d\mu^+\mu^-$ to avoid the
uncertainty while Refs. \cite{He2,Deshpande} separately estimate
the upper bounds of the coupling constants for $P^0 sd$ and
$P^0\mu^+\mu^-$. Borrowing the values from Refs.
\cite{He2,Deshpande}, in which $\lambda^p_{sd} $ and
$\lambda^{p}_{\mu\mu}$ are individually from the $K^0-\bar K^0$
mixing and muon magnetic dipole moment, we obtain
$\lambda_{PP}\equiv\lambda^{P}_{sd}\lambda^{P}_{\mu\mu}<5 \times
10^{-13}$, which is
consistent with the upper values of our windows.\\
3. We also explicitly consider $S\otimes P$ and $P\otimes S$ couplings
which are absent in Refs. \cite{He2,Deshpande}.\\
4. We note that our results of $\Gamma_{P^0}\sim 10^{-7}$ MeV and
$\tau_{P^0}\sim 10^{-14}$ sec are close to the upper and lower
bounds of $\Gamma_{P^0}<1.6\times 10^{-6}$ MeV and $\tau_{P^0}\geq
1.7\times 10^{-15}$ sec, given by Refs. \cite{Deshpande} and
\cite{sgoldtino2}, respectively.
\section*{Acknowledgements}
We would like to thank Prof. X. G. He for useful discussions. This work
is financially supported by the National Science Council of
Republic of China under the contract number NSC-94-2112-M-007-004.

\end{document}